\documentclass[preprint,superscriptaddress,preprintnumbers,amsmath,amssymb,prd]{revtex4}
\usepackage{graphicx}
\usepackage{amsmath}

\begin{document}

\title{Constraints on non-Newtonian gravity and axionlike particles from
measuring the Casimir force in nanometer separation range }

\author{
G.~L.~Klimchitskaya}
\affiliation{Central Astronomical Observatory at Pulkovo of the
Russian Academy of Sciences, Saint Petersburg,
196140, Russia}
\affiliation{Institute of Physics, Nanotechnology and
Telecommunications, Peter the Great Saint Petersburg
Polytechnic University, Saint Petersburg, 195251, Russia}
\author{
P.~Kuusk}
\affiliation{Institute of Physics, University of Tartu, W.~Ostwaldi 1, 
Tartu, 50411, Estonia}
\author{
V.~M.~Mostepanenko}
\affiliation{Central Astronomical Observatory at Pulkovo of the
Russian Academy of Sciences, Saint Petersburg,
196140, Russia}
\affiliation{Institute of Physics, Nanotechnology and
Telecommunications, Peter the Great Saint Petersburg
Polytechnic University, Saint Petersburg, 195251, Russia}
\affiliation{Kazan Federal University, Kazan, 420008, Russia}

\begin{abstract}
We obtain constraints on the Yukawa-type corrections to Newton's
gravitational law and on the coupling constant of axionlike
particles to nucleons following from the experiment on measuring
the Casimir force between an Au-coated microsphere and a silicon
carbide plate. For this purpose, both the Yukawa-type force and
the force due to two-axion exchange between nucleons are
calculated in the experimental configuration. In the interaction
range of Yukawa force exceeding 1~nm and for axion masses above
17.8~eV, the obtained constraints are much stronger than those
found previously from measuring the lateral Casimir force between
sinusoidally corrugated surfaces. These results are compared with
the results of other laboratory experiments on constraining
non-Newtonian gravity and axionlike particles in the relevant
interaction ranges.
\end{abstract}

\maketitle

\section{Introduction}

Many extensions of the Standard Model predict the existence of
light scalar and pseudoscalar particles which play an important
role in the current concepts of the world, but still remain to
be discovered experimentally. Among these particles, one could
mention arion, graviphoton, dilaton, goldstino, axion and other
axionlike particles, moduli, chameleon,
etc. \cite{1,2,3,4,5,6,7,8,9,10}.
An exchange of such particles between the constituents
of two macrobodies results in novel macroscopic forces in addition
to the familiar electromagnetic and gravitational interactions.
Thus, an exchange of one light scalar particle leads to the
Yukawa-type interaction potential \cite{11}, whereas the
interaction potential due to an exchange of one pseudoscalar
particle between two fermions depends on their spins \cite{12,13}.
It is interesting to note that a simultaneous exchange of two
pseudoscalar particles coupled to fermions by means of the
pseudoscalar Lagrangian results in the spin-independent
interaction potential \cite{12,13}. However, for fermions
possessing the pseudovector coupling to axions, the form of
effective potential due to two-axion exchange remains unknown
\cite {14}.

The possible existence of light scalar particles and associated
with them Yukawa-type forces between macroscopic bodies is of
great importance, because it is equivalent to the modification
of Newton's law of gravitation at short separations where this
fundamental law is not tested experimentally with sufficient
precision \cite{11}. We emphasize also that the Yukawa-type
corrections to Newtonian gravity have been predicted \cite{15,16}
in extradimensional physics assuming the low-energy compactification
scale of the order of 1~TeV \cite{17,18}. For these reasons, great
pains were taken to detect the Yukawa-type force which may manifest
itself in different experiments or at least to constrain its strength
and interaction range as severely as possible. The latter goal was
reached in the gravitational experiments of E\"{o}tvos and Cavendish
type \cite{19,20,21,22}, measurements of the normal and lateral
Casimir forces \cite{23,24,25,26,27,28,29,30,31,32,33}, and in
experiments on neutron scattering \cite{34,35,36,37,38,39}.

The Yukawa-type correction to the Newtonian gravitational potential 
appears also in the post-Newtonian approximation of the 
scalar-tensor theory of gravity \cite{39a,39b}. 
The constraints on this correction at the length scale of about 
astronomical unit were obtained using radio links with the Cassini 
spacecraft \cite{39b,39c}.

Various possibilities of observing the pseudoscalar axionlike
particles are equally important for both particle physics and
astrophysics and cosmology (see, e.g., reviews \cite{40,41,42}).
It has been known that axions help to resolve the problems of
lacking strong $CP$ violation in QCD and large electric dipole
moment of a neutron. These particles are also considered as the
most probable candidates for the constituents of dark matter.
For the mass-range of our interest, the most important searches
of axionlike particles are based on the Cavendish-type
experiments \cite{43,44}, experiments on measuring the
Casimir force \cite{45,46,47,48,49,50,51,52}, and measurement
of forces between protons in the beam of molecular
hydrogen \cite{53}, which allow an obtaining of competitive
laboratory constraints on the coupling constants of axions to
nucleons.

In this paper, we consider the hypothetical interactions
arising from the exchange of scalar and pseudoscalar particles
and obtain constraints on them which follow from recent measuring
the Casimir force between a gold-coated microsphere and a silicon
carbide plate \cite{54}. It has been known that in high-precision
measurements of the Casimir interaction made in vacuum the minimum
separation distance between the test bodies was more than 160~nm
\cite{26,27,55}. Rather competitive constraints on the Yukawa force
at a very short interaction range \cite{25} have been obtained,
however, from a less precise ambient measurement of the Casimir
force between two crossed cylinders performed at separations from
20 to 100~nm \cite{56}. Over a period of time, these constraints
were the strongest ones in the interaction range $\sim10~$nm. In the
more recent times, they were replaced by  stronger constraints
following from the experiment on neutron scattering \cite{37}
and from measurements of the lateral Casimir force between
sinusoidally corrugated surfaces \cite{29,58,59}. In the latter
case, the closest approach between the two interacting bodies was
only 22 and 23~nm in the first and second sets of measurements,
respectively \cite{58,59}. Just this has made possible obtaining
rather strong constraints on the Yukawa interaction in the range
of the order of 10~nm.

Measurements of the Casimir force in Ref.~\cite{54} have an
advantage that they were performed at even shorter separations
starting from 10~nm. The measurement data were compared with
theoretical predictions of the Lifshitz theory and the extent
of agreement between them has been quantified. This makes it
possible to find the constraints on any hypothetical interaction
which was not observed within the determined confidence
interval. One more advantage of measurements at very short
separations is that the comparison with theory in this case does
not suffer from an uncertainty as to the correct description of
free charge carriers in the Casimir effect. This unresolved
problem (the so-called Casimir puzzle \cite{55,57}) is of major
importance at separations exceeding 100~nm, but is immaterial
at shorter separations where the Casimir force is completely
determined by the bound (core) electrons.

We calculate the Yukawa-type force and the force due to
two-axion exchange in the experimental configuration of
Ref.~\cite{54} (note that an exchange of one pseudoscalar
particle does not contribute in this case because the test
bodies are not polarized). The constraints on the Yukawa
interaction constant and the coupling constant of axion to
nucleons are determined here at the 95\% confidence level.
According to our results, in the interaction range from 1 to
3.7~nm the obtained constraints on the Yukawa interaction
constant are up to a factor of $5\times 10^5$ stronger than those
obtained previously from measuring the lateral Casimir force
between sinusoidally corrugated surfaces \cite{29,58,59}. We
show also that the constraints on a coupling constant of
axionlike particles to nucleons, following from the experiment
of Ref.~\cite{54}, are stronger than those from measuring the
lateral Casimir force \cite{48,58,59} for the axion masses
satisfying a condition $m_a > 17.8$~eV. Within these interaction
ranges,  the stronger constraints on non-Newtonian gravity and
coupling constant on axions to nucleons have been found only from
the experiments on neutron scattering \cite{37,38} and hydrogen
beams \cite{53}, respectively.

The paper is organized as follows. In Sec.~II we present a few
necessary details of the experiment of Ref.~\cite{54} and the
expressions for hypothetical forces in the experimental
configuration. Section~III contains the derivation of
constraints on both the non-Newtonian gravity and axionlike
particles. In Sec.~IV the reader will find our conclusions
and the discussion of future prospects. We use the system of
units in which $\hbar=c=1$.

\section{The hypothetical forces in the Casimir experiment
{\protect{\\}}utilizing gold and silicon carbide test bodies}

In Ref.~\cite{54}, the Casimir force $F_C$ between a $N$-doped SiC plate
of $D=400~\mu$m thickness and an Au-coated borosilicate sphere of
$R=10~\mu$m radius was measured in dry N${}_2$ atmosphere 
by means of an atomic force microscope within
the separation region from $a=10~$nm to $a=200~$nm.
The thickness of an Au coating on the sphere was $d=100~$nm.
The optical properties of the plate have been characterized 
by means of ellipsometers in the range of 
wavelengths from 140~nm to $30~\mu$m \cite{60}.
Together with the optical data for Au measured earlier \cite{61},
this allowed calculation of the Casimir force using the Lifshitz theory
and comparison of theoretical results with the measurement data.

The most important for us are the force measurements at the shortest separations
(the first five force magnitudes from 3.5 to 1.6~nN were measured at separations
$a$ from 10.8 to 14.4~nm, respectively) and the measure of their agreement with theory. The dominant experimental error due to the error in determination
of the cantilever spring constant $\Delta F_C^{\rm expt}=0.35~$nN was independent
of separation. Taking into account that the experimental separations were
determined with an error $\Delta a\approx 1~$nm \cite{54}, the theoretical
forces calculated at the experimental separations were burdened by the error
$\Delta F_C^{\rm theor}=(2.48\Delta a/a)F_C^{\rm theor}$ which varies from 0.80
to 0.28~nN within the separation interval indicated above. Note that both
$\Delta F_C^{\rm theor}$ and $\Delta F_C^{\rm expt}$ have been determined
at the 67\% confidence level. Then, the half-width of the 95\% confidence
interval for the random quantity $F_C^{\rm theor}-F_C^{\rm expt}$ is
defined as \cite{57}
\begin{equation}
\Theta_{0.95}(a)=2\sqrt{\left[\Delta F_C^{\rm expt}\right]^2+
\left[\Delta F_C^{\rm theor}\right]^2}.
\label{eq1}
\end{equation}

It is easily seen that $\Theta_{0.95}$ decreases from 1.8 to 0.9~nN when the 
separation increases from 10.8 to 14.4~nm. One can conclude that the magnitude
of any hypothetical interaction, which was not observed in the experiment
of Ref.~\cite{54}, should satisfy the inequality
\begin{equation}
|F^{\rm hyp}(a)|<\Theta_{0.95}(a).
\label{eq2}
\end{equation}

Now we list  explicit expressions for the Yukawa-type force and for the force
arising due to two-axion exchange between nucleons of a sphere and a plate.
We start with the case of two atoms with masses $m_1$ and $m_2$ situated at
the points $\mbox{\boldmath$r$}_1$ of an Au layer coating the sphere and
$\mbox{\boldmath$r$}_2$ of a SiC plate. An exchange of a scalar particle
of mass $M=1/\lambda$ between these atoms, where $\lambda$ is the Compton
wavelength of a particle, results in the Yukawa potential which is
usually parametrized as a correction to Newton's gravitational potential
\cite{11}
\begin{equation}
V_{\rm Yu}(r_{12})=-\frac{Gm_1m_2}{r_{12}}\alpha e^{-r_{12}/\lambda}.
\label{eq3}
\end{equation}
\noindent
Here, $G$ is the Newtonian gravitational constant, $\alpha$ is the dimensionless
constant of Yukawa interaction, and
$r_{12}=|\mbox{\boldmath$r$}_1-\mbox{\boldmath$r$}_2|$.

By integrating Eq.~(\ref{eq3}) over the volumes of a SiC plate
(which can be safely
considered as having an infinitely large area) and
of an Au spherical envelope, and calculating the negative derivative with
respect to the separation $a$, one arrives to the following expression for
the force (see Refs.~\cite{62,63} for details):
\begin{eqnarray}
&&
F_{\rm Yu}(a)=-4\pi^2G\alpha\rho_1\rho_2\lambda^3e^{-a/\lambda}
\nonumber \\
&&~~~~~~~~~\times
\left[\Phi(R,\lambda)-\Phi(R-d,\lambda)e^{-d/\lambda}\right].
\label{eq4}
\end{eqnarray}
\noindent
Here, $\rho_1=19.28~\mbox{g/cm}^3$ and $\rho_2=3.21~\mbox{g/cm}^3$ are the
densities of Au and SiC, respectively, and the function $\Phi$ is defined as
\begin{equation}
\Phi(r,\lambda)=r-\lambda+(r+\lambda)e^{-2r/\lambda}.
\label{eq5}
\end{equation}

Note that in Eq.~(\ref{eq4}) we disregarded by the contribution of a borosilicate
sphere core to the hypothetical force. Its account could lead to only minor
increase of $|F_{\rm Yu}|$ and, thus, to slightly stronger constraints on
$\alpha$. We have also considered the SiC plate as an infinitely thick because in
the region of $\lambda$ under consideration
$\exp(-D/\lambda)=0$ at high precision.

We are coming now to two-axion exchange between the nucleons (protons and
neutrons), which belong to an Au layer coating the sphere and a SiC plate,
situated at the points $\mbox{\boldmath$r$}_1$ and $\mbox{\boldmath$r$}_2$,
respectively. Assuming the pseudoscalar coupling between an axion and a
nucleon, one obtains the following interaction potential \cite{12,13}
\begin{equation}
V_a(r_{12})=-\frac{g_{an}^4}{32\pi^3m^2}\,\frac{m_a}{r_{12}^2}\,
{K}_1(2m_ar_{12}),
\label{eq6}
\end{equation}
\noindent
where $g_{an}$ is the dimensionless interaction constant of an axion to a
nucleon (it is assumed that the interaction constants of an axion to a
proton and a neutron are similar \cite{13}), $m_a$ is the axion mass, $m$
is the mean nucleon mass, and ${K}_1(z)$ is the modified Bessel
function of the second kind.

Integrating Eq.~(\ref{eq6}) over the volumes of an Au spherical layer
and a SiC plate and performing the negative differentiation with respect to
$a$, one obtains the following result (see Refs.~\cite{49,52} for the
details of calculation):
\begin{eqnarray}
&&
F_a(a)=-\frac{\pi}{2m_am^2m_{\rm H}^2} C_1C_2\!\int_1^{\infty}\!\!\!du
\frac{\sqrt{u^2-1}}{u^3}e^{-2m_aua}
\nonumber \\
&&~~~~~\times
\left[\chi(R,m_au)- \chi(R-d,m_au)e^{-2m_aud}\right].
\label{eq7}
\end{eqnarray}
\noindent
Here, $m_{\rm H}$ is the mass of atomic hydrogen,
\begin{equation}
C_i=\rho_i\frac{g_{an}^2}{4\pi}\frac{Z_i+N_i}{\mu_i},
\label{eq8}
\end{equation}
\noindent
where $i=1,\,2$ for Au and SiC, respectively, $Z_i$ and $N_i$ are the
number of protons and the mean number of neutrons, and $\mu_i=m_i/m_{\rm H}$,
$m_i$ being the mean atomic (molecular) mass. According to information
presented in Ref.~\cite{11}, one has
\begin{eqnarray}
&&
\frac{Z_1}{\mu_1}=0.40422,{\ } \quad \frac{N_1}{\mu_1}=0.60378,
\nonumber \\
&&
\frac{Z_2}{\mu_2}=0.502701, \quad \frac{N_2}{\mu_2}=0.505708.
\label{eq9}
\end{eqnarray}
\noindent
Finally, the function $\chi$ in Eq.~(\ref{eq7}) is  defined as
\begin{equation}
\chi(r,z)=r-\frac{1}{2z}+\left(r+\frac{1}{2z}\right)e^{-2rz}.
\label{eq10}
\end{equation}

In the next section, the above expressions are used for constraining the
parameters of Yukawa-type forces and axionlike particles from the measurement
results of Ref.~\cite{54}.

\section{Constraints on non-Newtonian gravity and axionlike particles}

To obtain constraints on the hypothetical forces of Yukawa-type,
we have substituted the force $F_{Yu}$ defined in
Eqs.~(\ref{eq4}) and (\ref{eq5}) in Eq.~(\ref{eq2}) in place of
$F^{\rm hyp}$. The obtained inequality was analyzed in the region
of $a$ from 10.8 to 14.4~nm. The constraints on $\alpha$ following
from this inequality as a function of $\lambda$ are shown by the
line 2 in Fig.~1. In the same figure, the constraints on $\alpha$
found \cite{29} from experiments on measuring the lateral Casimir
force between sinusoidally corrugated surfaces \cite{58,59} are
shown by the line 1.

As is seen in Fig.~1, the constraints of line 2 are stronger than
those of line 1 within the range of $\lambda$ from 1 to 3.7 nm.
The largest strengthening by the factor of $495000\approx 5\times10^5$
is reached at $\lambda=1~$nm. These are the strongest constraints
for small $\lambda$ obtained to date from measuring
the Casimir force. For even smaller $\lambda$, where the discrete
structure of matter should be taken into account, the strongest
constraints on the Yukawa-type interaction are obtained from the
atomic physics \cite{64}.

Within the interaction range $\lambda\geqslant 1~$nm the constraints
of line 2 are not the strongest ones. The strongest constraints on
$\alpha$ obtained in this range so far follow from the experiments
on neutron scattering. They are shown by the lines 3 \cite{37} and
4 \cite{38}. The neutron scattering provides the strongest
constraints on $\alpha$ within the interaction range up to
$\lambda=10~$nm. For larger $\lambda$ the strongest constraints on
$\alpha$ are shown by the line 5 obtained \cite{30} from measuring
the normal Casimir force between sinusoidally corrugated
curfaces \cite{65,66}.

Note that the lines 1--4 have been obtained at the 95\% confidence
level, whereas the line 5 was found at the 67\% confidence level
\cite{30,65,66}. For even longer interactions 31~nm$<\lambda<8~\mu$m
the strongest constraints on $\alpha$ were obtained in the
Casimir-less experiment of Ref.~\cite{31} (line 6), and in the
interaction range $\lambda>8~\mu$m from the experiments of
Cavendish-type \cite{19,20,21,22} (line 7).
These two lines were obtained at the
95\% confidence level. Recall that the regions of
($\lambda,\,\alpha$)-plane situated above each line in Fig.~1 are
excluded by the results
of respective experiments, whereas the regions below each line are
allowed.

As mentioned in Sec.~I, rather strong constraints on $\alpha$ for
small $\lambda$ were obtained \cite{25} from measuring the Casimir
force between two crossed cylinders \cite{56}. These constraints,
however, are not shown in Fig.~1 because they cannot be characterized
by some definite confidence level. The point is that the Au films in
the experiment of Ref.~\cite{56} were fixed to silica cylinders by means
of a soft glue. Furthermore, the presence of a hydrocarbon organic layer
on the interacting surfaces did not allow to quantify the measure of
agreement between experiment and theory \cite{57}.

We now turn our attention to the derivation of constraints on axionlike
particles. For this purpose, in place of $F^{\rm hyp}$ in Eq.~(\ref{eq2}) we
substitute the force from Eqs.~(\ref{eq7})--(\ref{eq10}) arising due to
two-axion exchange between the nucleons of an Au spherical layer and a
SiC plate. The numerical analysis of the obtained inequality performed
similar to the case of Yukawa interaction leads to the line 1 in
Fig.~2 demonstrating the obtained constraints on $g_{an}^2/(4\pi)$
as a function of the axion mass $m_a$.

In the same figure, the constraints of line 2 were obtained in
Ref.~\cite{48} from measuring the lateral Casimir force between
sinusoidally corrugated surfaces \cite{58,59}, and the constraints
of line 3 were found in Ref.~\cite{47} from measuring the effective
Casimir pressure \cite{26,27}. Furthermore, the lines 5 and 6 in Fig.~2
show the constraints derived \cite{49} from the Casimir-less experiment
of Ref.~\cite{31} and from the Cavendish-type experiments \cite{43,44},
respectively.

As is seen in Fig.~2, in the region of axion masses $m_a>17.8~$eV the
constraints of line 1 are stronger than all the above listed
constraints obtained from the Casimir effect in the past. It should
be noted, however, that the line 4 obtained from the measure of
agreement between experiment and theory for forces acting between
protons in the beam of molecular hydrogen \cite{53} (see also
Ref.~\cite{67}) presents much stronger constraints for axion masses
$m_a>0.5~$eV. For heavier axions with $m_a>200~$eV, further strengthening
of constraints of Ref.~\cite{53} on $g_{an}$ was obtained in Ref.~\cite{67}
by comparing the nuclear magnetic resonance experiment and theory
for nucleons in deuterated molecular hydrogen. As to the lightest
axions with masses $m_a<1~\mu$eV, the strongest constraints on their
coupling constant to nucleons were obtained from the magnetometer
measurements using spin-polarized K and ${}^3$He atoms \cite{68} (see the
line 7). Similar to Fig.~1, the regions of the plane above each line are
excluded and below each line are allowed by respective experiments.
All the constraints under consideration have been obtained at the
95\% confidence level.

\section{Conclusions and discussion}

In this paper, we have obtained constraints on the Yukawa-type
corrections to Newtonian gravitational law and the coupling
constant of axionlike particles to nucleons which follow from
measuring the Casimir force between an Au-coated microsphere
and a silicon carbide plate. An advantage of this experiment
operated in dry N${}_2$ atmosphere, as compared to more precise
experiments operated in high vacuum, is that the force was
measured down to 10~nm separation between the sphere and the
plate. In spite of rather large experimental and theoretical
errors, this allowed to obtain stronger constraints on the
hypothetical interactions at sufficiently short distance range 
of the order of 1~nm (or for relatively large axion masses of
more than 17.8~eV) than have been obtained previously from
more precise measurements of the Casimir force performed at
larger separation distances.

Specifically, in the interaction range $\lambda$ from 1 to 3.7~nm
the constraints on the Yukawa interaction following from the
experiment using a SiC plate are by up to a factor $5\times10^5$
stronger than those found previously from measuring the lateral
Casimir force between sinusoidally corrugated surfaces. The
constraints on the coupling constants of axionlike particles to
nucleons following from the experiment using a SiC plate are
stronger than those obtained from measuring the lateral Casimir
force for the axion masses satisfying a condition $m_a > 17.8$~eV.
It should be noted, however, that in these regions of relatively
small $\lambda$ and large $m_a$ the strongest current constraints
on non-Newtonian gravity and axion to nucleon coupling constant
are provided not by the Casimir effect, but by the neutron
scattering and hydrogen beams, respectively.

It is pertinent to note that experiments on measuring the Casimir
force have potentialities for obtaining stronger constraints on
both non-Newtonian gravity and axionlike particles. Thus, as
shown in Ref.~\cite{52}, the constraints of Figs.~1 and 2
following from measuring the lateral and normal Casimir force
between corrugated surfaces can be strengthened significantly
by modifying the parameters of corrugations. Similar effect
can be reached \cite{52} by choosing thicker sectors of the
patterned plate used in the Casimir-less experiment of
Ref.~\cite{31}. One could mention also the recently proposed
Casimir experiments at larger separation distances between the
test bodies \cite{69,70,71,72}. Several other very recent
suggestions for searching the non-Newtonian gravity are based
on measuring the Casimir-Polder interaction in the presence of
a metallic shield \cite{73}, the normal mode  splitting in the
optomechanical cavity \cite{74}, and on precision spectroscopy
of weakly bound molecules \cite{75}. Taken together, these
approaches should lead to further strengthening of the
constraints on non-Newtonian gravity and axionlike particles.

\section*{Acknowledgments}
The work of G.L.K.~and V.M.M.~was partially supported by the
Peter the Great Saint Petersburg Polytechnic University in
the framework of the Program ``5--100--2020".
The work of P.K.\ was supported by the European Union through the 
European Regional Development Fund CoE program grant TK133 
``The Dark Side of the Universe".
V.M.M.~was partially funded by the Russian Foundation for Basic
Research, Grant No. 19-02-00453 A. His work was also partially
supported by the Russian Government Program of Competitive Growth
of Kazan Federal University.
G.L.K.~and V.M.M.~are grateful to the Institute
of Physics of the University of Tartu, where this
work was performed, for kind hospitality.

\begin{figure}[b]
\vspace*{-6cm}
\centerline{\hspace*{0.3cm}
\includegraphics{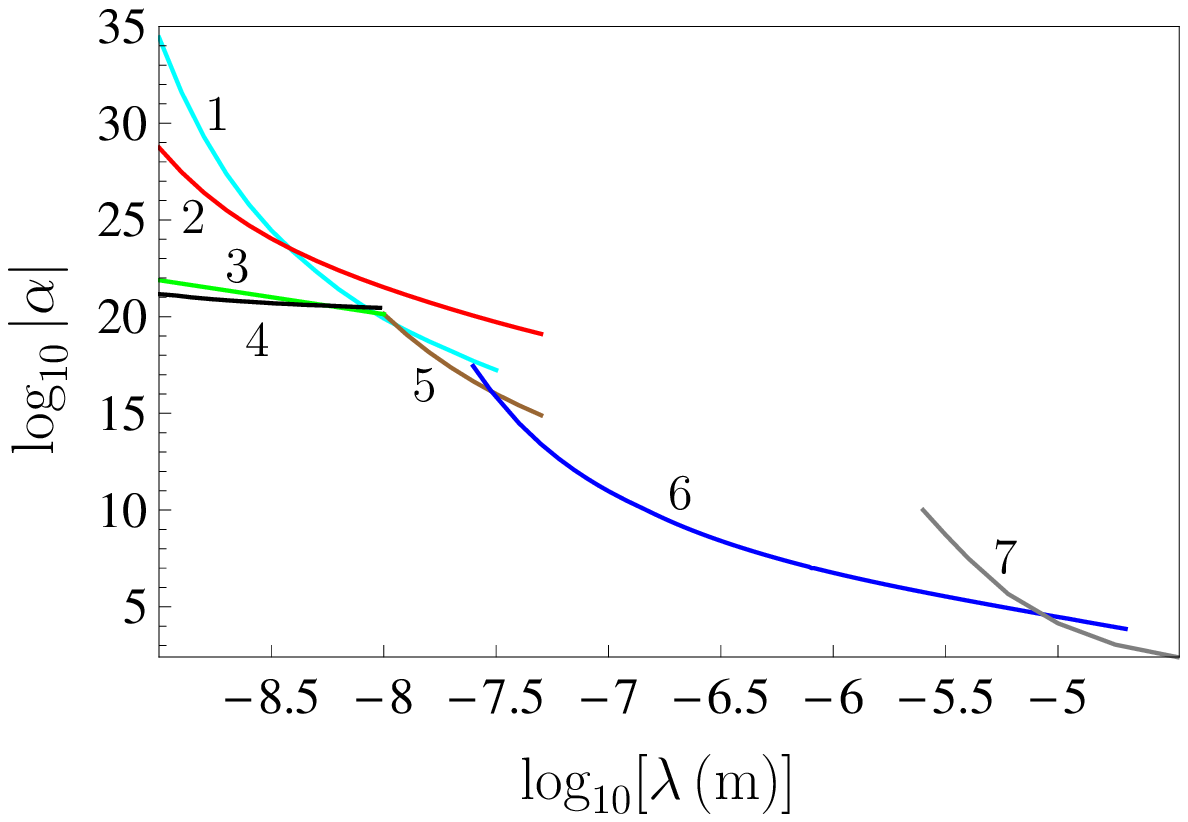}
}
\vspace*{-11.cm}
\caption{\label{fg1}
Constraints on the strength of a Yukawa-type correction to
Newton's gravitational law obtained from measuring the lateral
Casimir force between corrugated surfaces (line 1), in this work
from the experiment using a silicon carbide plate (line 2), from
the experiments on neutron scattering (lines 3 and 4), from
measuring the normal Casimir force between corrugated surfaces
(line 5), from the Casimir-less experiment (line 6), and from
the gravitational experiments (line 7) are shown as the functions
of the interaction range (see the text for further discussion).
The regions of the $(\lambda,\,\alpha)$-plane above each line are
excluded and below each line are allowed by the experimental results.
}
\end{figure}
\begin{figure}[b]
\vspace*{2cm}
\centerline{\hspace*{0.3cm}
\includegraphics{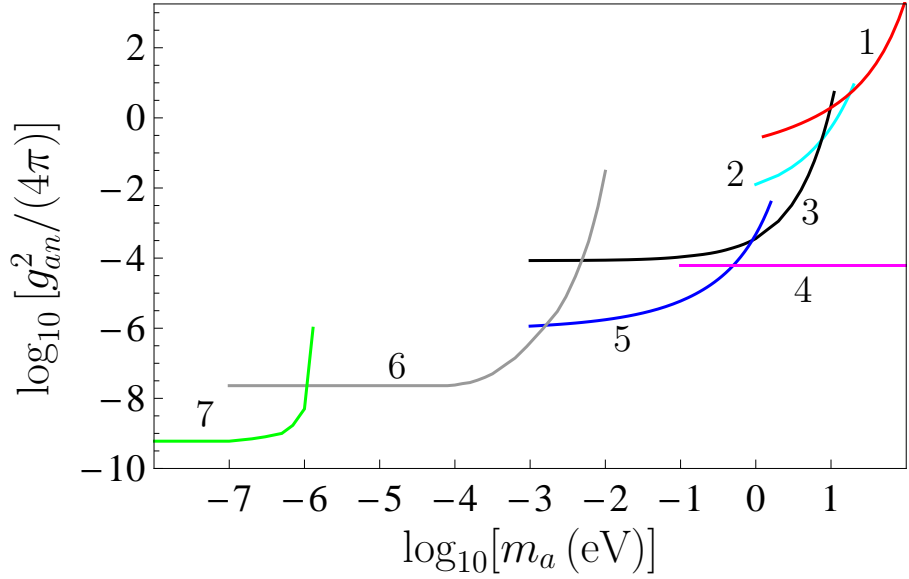}
}
\vspace*{-19.cm}
\caption{\label{fg2}
Constraints on the coupling constant of axionlike particles to
nucleons obtained in this work from the experiment using a
silicon carbide plate (line 1), from measuring the lateral
Casimir force between corrugated surfaces (line 2), from
measuring the effective Casimir pressure (line 3), from
the experiment using a beam of molecular hydrogen (line 4),
from the Casimir-less experiment (line 5), from
the gravitational experiments (line 6), and from the
magnitometer measurements (line 7) are shown as the functions
of the axion mass (see the text for further discussion).
The regions of the $[m_a,\,g_{an}^2/(4\pi)]$-plane above each line are
excluded and below each line are allowed by the experimental results.
}
\end{figure}
\end{document}